\documentclass[twocolumn]{aastex63}
\usepackage{amsmath}

\newcommand{\cii}{[C\,{\sc ii}]}
\newcommand{\ci}{[C\,{\sc i}]}
\newcommand{\oi}{[O\,{\sc i}]}

\submitjournal{ApJ}

\shorttitle{Sub-kpc structure of a lensed quasar at $z=6.5$}
\shortauthors{Yue et al.}
\graphicspath{{./}{figures/}}
\begin{document}

\title{ALMA Observations of the Sub-kpc Structure of the Host Galaxy of a $z=6.5$ Lensed Quasar:
A Rotationally-Supported Hyper-Starburst System at the Epoch of Reionization}

\correspondingauthor{Minghao Yue}
\email{yuemh@email.arizona.edu}

\author[0000-0002-5367-8021]{Minghao Yue}
\affiliation{Steward Observatory, University of Arizona, 933 North Cherry Avenue, Tucson, AZ 85721, USA}

\author[0000-0001-5287-4242]{Jinyi Yang}
\altaffiliation{Strittmatter Fellow}
\affil{Steward Observatory, University of Arizona, 933 North Cherry Avenue, Tucson, AZ 85721, USA}

\author[0000-0003-3310-0131]{Xiaohui Fan}
\affiliation{Steward Observatory, University of Arizona, 933 North Cherry Avenue, Tucson, AZ 85721, USA}

\author[0000-0002-7633-431X]{Feige Wang}
\altaffiliation{NHFP Hubble Fellow}
\affiliation{Steward Observatory, University of Arizona, 933 North Cherry Avenue, Tucson, AZ 85721, USA}

\author[0000-0003-3256-5615]{Justin~Spilker}
\altaffiliation{NHFP Hubble Fellow}
\affiliation{Department of Astronomy, University of Texas at Austin, 2515 Speedway, Stop C1400, Austin, TX 78712, USA}

\author[0000-0001-8471-6679]{Iskren Y. Georgiev}
\affiliation{Max-Planck Institute for Astronomy, K{\"o}nigstuhl 17, D-69117 Heidelberg, Germany}

\author[0000-0001-6812-2467]{Charles R. Keeton}
\affiliation{Department of Physics and Astronomy, Rutgers University, Piscataway, NJ 08854, USA}

\author[0000-0002-4208-3532]{Katrina C. Litke}
\affiliation{Steward Observatory, University of Arizona, 933 North Cherry Avenue, Tucson, AZ 85721, USA}

\author[0000-0002-2367-1080]{Daniel P. Marrone}
\affiliation{Steward Observatory, University of Arizona, 933 North Cherry Avenue, Tucson, AZ 85721, USA}

\author[0000-0003-4793-7880]{Fabian Walter}
\affiliation{Max-Planck Institute for Astronomy, K{\"o}nigstuhl 17, D-69117 Heidelberg, Germany}

\author[0000-0003-4956-5742]{Ran Wang}
\affiliation{Kavli Institute for Astronomy and Astrophysics, Peking University, Beijing 100871, China}

\author[0000-0002-7350-6913]{Xue-Bing Wu}
\affiliation{Kavli Institute for Astronomy and Astrophysics, Peking University, Beijing 100871, China}

\author[0000-0001-9024-8322]{Bram P.\ Venemans}
\affiliation{Max-Planck Institute for Astronomy, K{\"o}nigstuhl 17, D-69117 Heidelberg, Germany}

\author[0000-0001-6047-8469]{Ann Zabludoff}
\affiliation{Steward Observatory, University of Arizona, 933 North Cherry Avenue, Tucson, AZ 85721, USA}

%\collaboration{1}{(AAS Journals Data Scientists collaboration)}

%\author{Butler Burton}
%\affiliation{Leiden University}
%\affiliation{AAS Journals Associate Editor-in-Chief}
%\nocollaboration{1}

%\altaffiliation{AASTeX v6+ programmer}

%\collaboration{1}{(LaTeX collaboration)}

%\affiliation{AAS Director of Publishing}
%\affiliation{American Astronomical Society \\
%1667 K Street NW, Suite 800 \\

%\author{Scott Chernoff}
%\affiliation{IOP Publishing, Washington, DC 20005}




\begin{abstract}
We report ALMA observations of the dust continuum and {\cii} emission of the host galaxy of J0439+1634, a gravitationally lensed quasar at $z=6.5$. Gravitational lensing boosts the source-plane resolution to $\sim0\farcs15$ $(\sim0.8\text{ kpc})$. The lensing model derived from the ALMA data is consistent with the fiducial model in \citet{fan19} based on {\it HST} imaging. The host galaxy of J0439+1634 can be well-fitted by a S\'ersic profile consistent with an exponential disk, both in the far-infrared (FIR) continuum and the {\cii} emission. The overall magnification is $4.53\pm0.05$ for the continuum and $3.44\pm0.05$ for the {\cii} line. The host galaxy of J0439+1634 is a compact ultra-luminous infrared galaxy, with a total star formation rate (SFR) of $1.56\times10^{3}M_\odot/\text{year}$ after correcting for lensing and an effective radius of $0.74$ kpc. The resolved regions in J0439+1634 follow the ``{\cii} deficit," where the {\cii}-to-FIR ratio decreases with FIR surface brightness. The reconstructed velocity field of J0439+1634 appears to be rotation-like. The maximum line-of-sight rotation velocity of 130 km/s at a radius of 2 kpc. However, our data cannot be fit by an axisymmetric thin rotating disk, and the inclination of the rotation axis, $i$, remains unconstrained. We estimate the dynamical mass of the host galaxy to be $7.9\sin^{-2}(i)\times10^{9}M_\odot$. J0439+1634 is likely to have a high gas-mass fraction and an oversized SMBH compared to local relations. The SFR of J0439+1634 reaches the maximum possible values, and the SFR surface density is close to the highest value seen in any star-forming galaxy currently known in the universe.
\end{abstract}



\section{Introduction} \label{sec:intro}

%Luminous quasars at $z\gtrsim6$ are powerful (but rare) probes of 
%supermassive black holes (SMBHs) at high-redshift.
%SMBHs evolve 

In the past two decades, more than 200 quasars at 
$z>6$ have been discovered
\citep[e.g.,][]{matsuoka16, matsuoka18a, matsuoka2018b, matsuoka19, venemans13, venemans15, yang19a, yang20, wang17, wang19a, jiang16, banados16, banados18}.
%These quasars form a comprehensive sample which covers a wide range
%of luminosity and supermassive black hole (SMBH) mass.
Studies of the quasar host galaxies provide crucial knowledge
about the co-evolution of supermassive black holes (SMBHs)
with their host galaxies and environment in the early universe.
Detecting the host galaxies of high-redshift quasars is challenging
at rest-frame ultraviolet to near-infrared wavelengths,
where the emission from the central quasar overwhelms
the host galaxy \citep[e.g.,][]{mechtley12, marshall20}. As such,
information about quasar host galaxies is mostly
from the far-infrared (FIR) and sub-millimeter (sub-mm) regime
\citep[e.g.,][]{riechers09,wangran08,wangran10,venemans12}.
The dust continuum and atomic and molecular emission lines 
(for example, the {\cii} fine structure line and CO rotational lines)
contain a wealth of information about the interstellar medium
(ISM), including the dust mass and temperature
\citep[e.g.,][]{beelen06, Schreiber18}, 
the atomic and molecular gas mass \citep[e.g.,][]{weib05, bolatto13},
and the gas-phase metallicity \citep[e.g.,][]{rigopoulou18}.
Spatially resolved line emission also directly probes the gas-phase kinematics of quasar host galaxies and provides 
 the only current way to measure their dynamical masses 
\citep[e.g.,][]{walter09}.
The total-infrared (TIR) luminosity 
is  widely used to estimate the star formation rate (SFR) 
\citep[e.g.,][]{murphy11}, assuming that the cool dust in the quasar host is dominantly heated by star formation
\citep[e.g.,][]{beelen06, leipski14}.

With unprecedented sensitivity and resolving power,
the Atacama Large Millimeter/Submillimeter Array (ALMA)
has greatly improved our understanding of high-redshift quasars.
%At $z\gtrsim6$, quasar host galaxies usually have sizes of
%$0\farcs5\sim1\farcs0$ \citep[e.g.,][]{decarli18, wang19b},
%though they can be as compact as $\sim0\farcs2$ \citep[e.g.,][]{venemans17}.
%Correspondingly, most ALMA observations of high-redshift quasars
% use beam sizes of $\gtrsim 0\farcs3$,
%which can resolve quasar hosts into several beams
%\citep[e.g.,][]{venemans18, izumi19,Carniani19}.
To date, several tens of quasars at $z>6$ have been observed by ALMA,
including about 15 at $z>6.5$.
These observations led to an overall picture 
of the high-redshift quasar population:
most of the high-redshift quasars are hosted by infrared-luminous, gas rich
galaxies \citep[e.g.,][]{decarli18,venemans18, wang19b,shao19},
indicating active star formation 
$(\text{SFR}\gtrsim10^2M_\odot\text{ yr}^{-1})$.
The kinematics of the bright {\cii} emission line
 constrains the dynamical masses of quasar host galaxies.
Compared to the local $M_\text{BH}-M_\text{host}$ relation 
\citep[e.g.,][]{kh13}, SMBHs in quasars at $z\gtrsim6$ are oversized
\citep[e.g.,][]{venemans16,decarli18,wangran19}.
While SMBHs might grow earlier than their hosts at 
high redshift, this difference may be a result of 
selection effects, i.e., current quasar surveys
are biased toward luminous quasars, which have massive SMBHs
\citep[e.g.,][]{willott15,izumi19}.

At $z>6$, quasar host galaxies usually have sizes of
$\sim2 - 4\text{ kpc}$ \citep[e.g.,][]{decarli18},
although they can be as compact as $\sim1\text{ kpc}$
\citep[e.g.,][]{venemans17}.
Most ALMA observations of high-redshift quasars
use beam sizes of $\gtrsim 0\farcs3$,
which marginally resolve these quasar host galaxies.
These hosts have a variety of morphologies,
ranging from a regular Gaussian profile
\citep[e.g.,][]{shao17,venemans18} to highly irregular, 
indicating an on-going merging system 
\citep[e.g.,][]{banados19, neeleman19}.
In a recent study, \citet{venemans19} reported
400-pc resolution imaging of a quasar host galaxy at redshift 6.6,
which shows complex structures of dust continuum and {\cii} emission,
including cavities with sizes of $\sim 0.5 \text{ kpc}$.
The authors propose that these cavities might be relevant to 
the energy output of the central active galactic nucleus (AGN).
Sub-kpc resolution is thus necessary to the investigation of the 
structures in high-redshift quasars and to the understanding of
SMBH-host coevolution.

Gravitational lensing acts as a natural telescope,
 significantly enhancing the angular resolution and the sensitivity
of observations \citep[e.g.,][]{hezaveh16,litke19,inoue20,cheng20,spilker20}.
In \citet{fan19}, we reported the discovery of a gravitationally lensed
quasar at $z=6.51$, J043947.08+163415.7 (hereafter J0439+1634).
High-resolution images taken by \textit{Hubble Space Telescope} ({\it HST})
reveal the multiple images of the quasar generated by gravitational lensing.
The lensing model based on {\it HST} data suggests that 
J0439+1634 is a naked-cusp lens with three images, 
with a total magnification of $51.3\pm1.4$. 
J0439+1634 is the only known lensed quasar at $z>5$ to date
and provides an excellent chance to study a high-redshift quasar
in enhanced spatial resolution due to its large lensing magnification. 
%\citet{yang19} reported strong FIR continuum and {\cii} emission for J0439+1634,
%yet 

Here we report the sub-mm continuum and {\cii} $158~\mu\text{m}$ emission line
of J0439+1634 observed by ALMA at a resolution of $\sim 0\farcs3$.
With the help of lensing, we reach a physical resolution of 
$\sim0.8 \text{ kpc}$.
%when averaged over the entire galaxy, and
%higher resolution in the nuclear region of the galaxy. 
%($\sim0.2 \text{ kpc}$ within the central 0.2 kpc) 
We describe our data in Section \ref{sec:data}.
In Section \ref{sec:lensing}, we describe the measurement
of the dust continuum and {\cii} emission line, including the lens modeling
and the reconstruction of the velocity field.
We present the physical properties of J0439+1634 in Section \ref{sec:physical}
and discuss their implications for the evolutionary state of the quasar host galaxy in Section \ref{sec:discuss}.
We summarize this paper in Section \ref{sec:conclude}.
Throughout this paper, we use a $\Lambda$CDM universe 
with $H_0=70 \text{ km s}^{-1}\text{Mpc}^{-1}$,
$\Omega_\text{M}=0.3$, and $\Omega_\Lambda=0.7$.

\section{Data} \label{sec:data}

J0439+1634 was observed in ALMA Band 6 under configuration C43-5 in October 2018.
The configuration contains 48 12-m antennas, which has a
 maximum baseline of 1.24 km.
We tuned the four 1.875 GHz-wide spectral windows (SPWs) at 
238.593 GHz, 236.718 GHz, 252.206 GHz, and 253.894 GHz
with channel widths of 
15.625 MHz, 15.625 MHz, 7.8125 MHz, and 7.8125 MHz, respectively.
The {\cii} emission line falls in the third SPW.
The on-source exposure time is 99 minutes.
We use J0510+1800 as the bandpass calibrator and
J0440+1437 as the phase calibrator.
The C43-5 observations are a part of Program 2018.1.00566.S,
which aims at a mapping the dust continuum and {\cii} emission line
to a spatial resolution of $0\farcs03$.
The high-resolution observation with configuration C43-8
is not completed at the time of this paper's writing.
%{\bf mention the whole scope of the program includes high resolution configuation, which is yet to be completed at the time of this writing.}
\begin{figure*}
    \centering
    \includegraphics[width=0.4\textwidth,trim=0cm 0cm 0cm 0cm,clip]{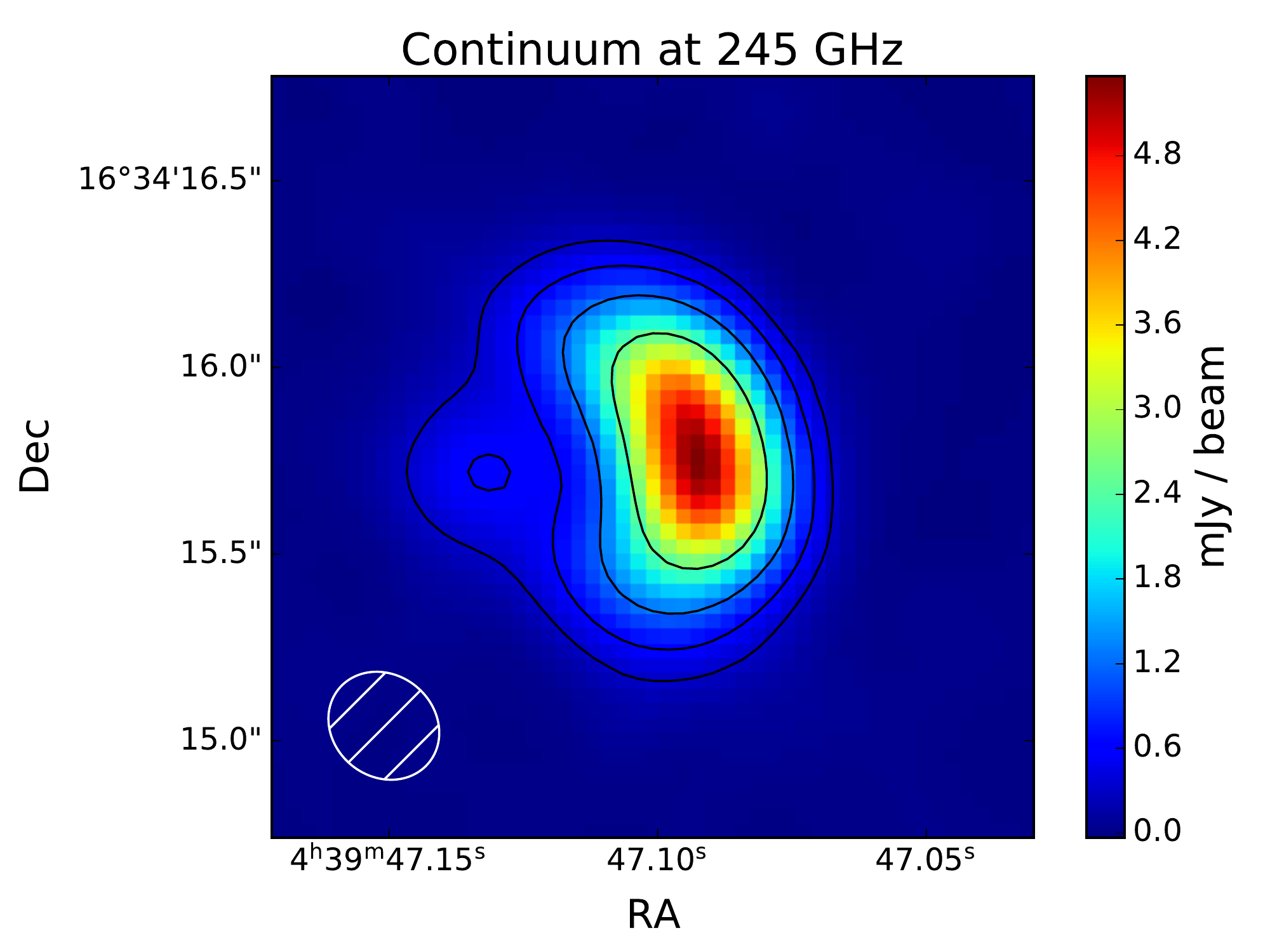}
    \includegraphics[width=0.4\textwidth,trim=0cm 0cm 0cm 0cm,clip]{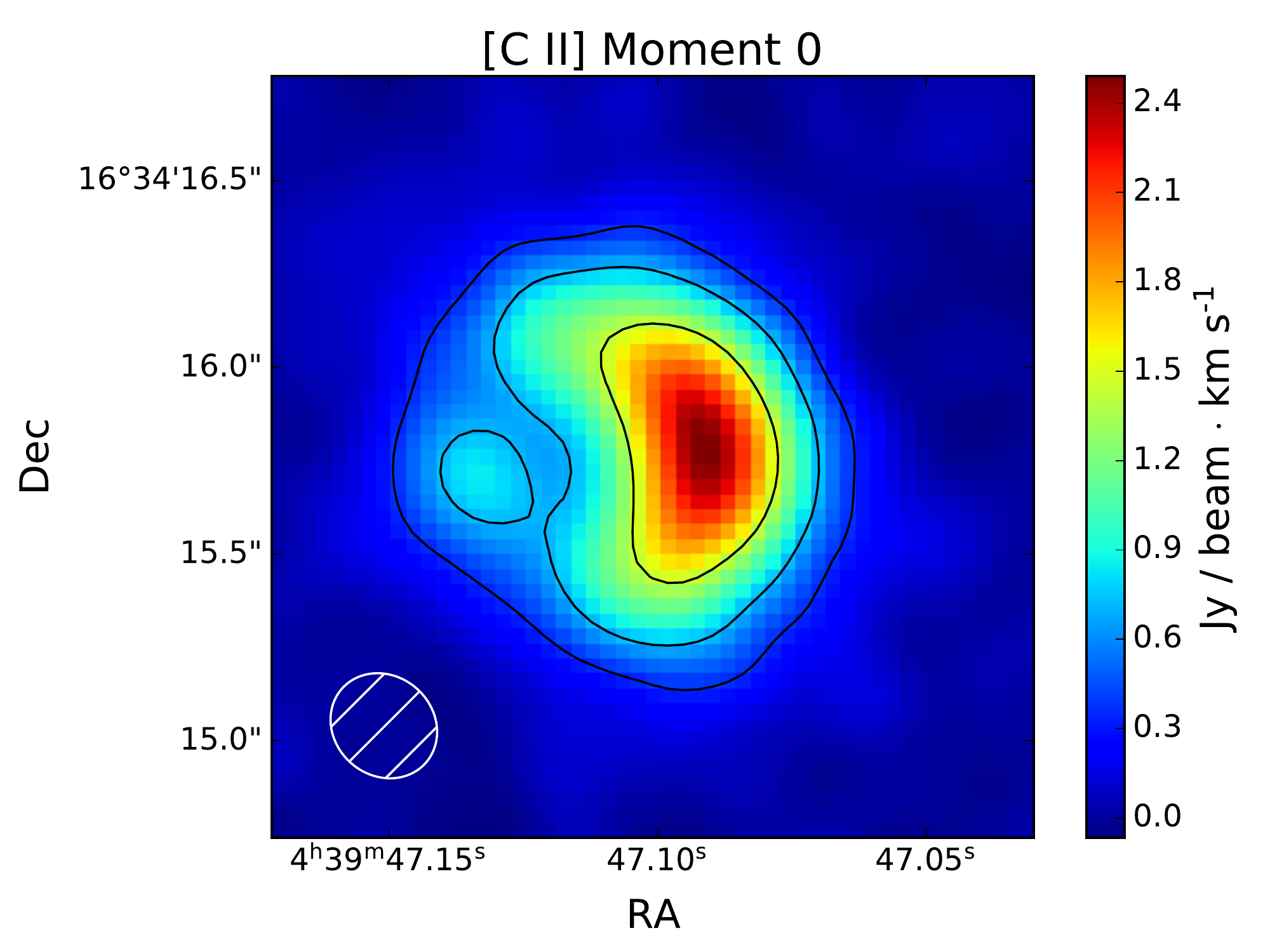}
    \includegraphics[width=0.4\textwidth,trim=0cm 0cm 0cm 0cm,clip]{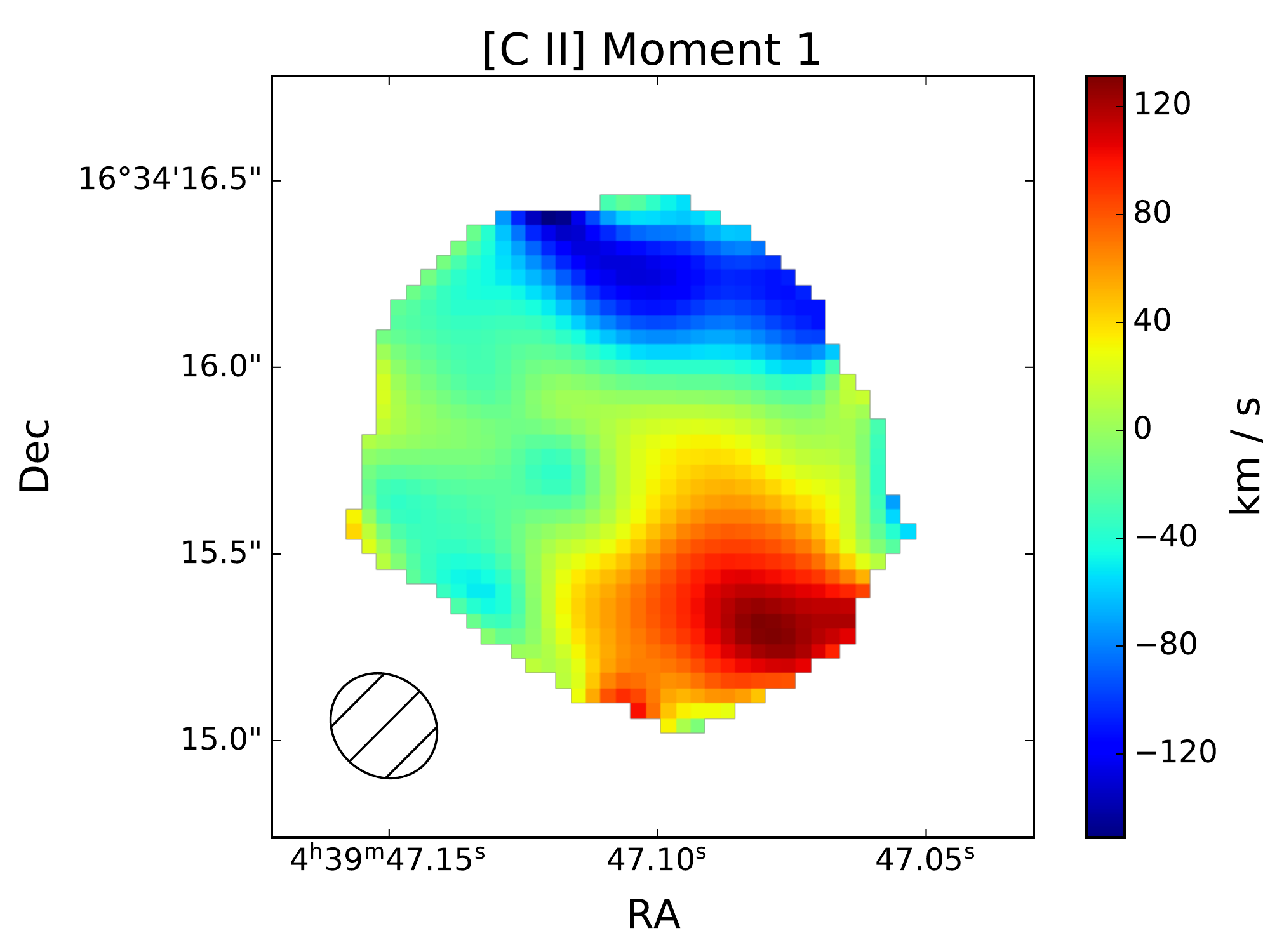}
    \includegraphics[width=0.4\textwidth,trim=0cm 0cm 0cm 0cm,clip]{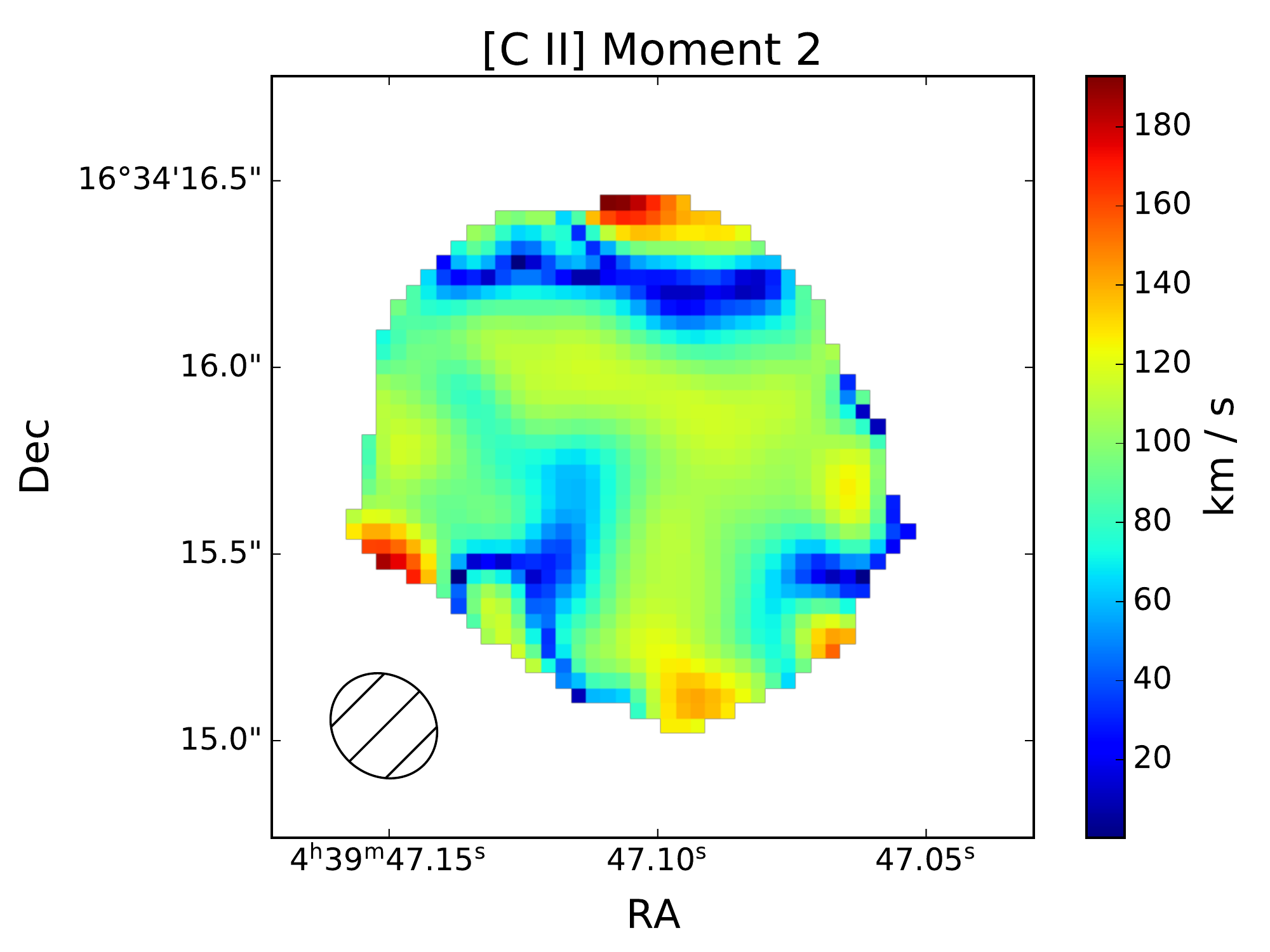}
    \caption{The clean image and moments of J0439+1634 observed with ALMA. 
    {\it Upper Left}: the continuum; 
    {\it Upper Right}: the zeroth moment of {\cii} emission; 
    {\it Lower Left} and {\it Lower Right}: the first and second moments
    of the {\cii} emission.
    The data are cleaned using Briggs weighting with $\text{robust}=0.5$.
    In the first and second moment maps, we only show pixels that have 
    integrated flux signal-to-noise ratio larger than 10.
    Contours in the continuum map are at $20\sigma$, $40\sigma$,
    $80\sigma$ and $160\sigma$ levels, where $\sigma$ is the flux error per beam
    estimated using an annulus with $1''<r<2''$.
    Similarly, Contours in the {\cii} moment 0 map are at $10\sigma$, $20\sigma$ and
    $40\sigma$ levels. The moment 1 map shows rotation-like ordered motion.}
    \label{fig:clean}
\end{figure*}

We reduce the ALMA data using the Common Astronomy Software Applications (CASA)
version 5.6.1 \citep{casa}.
We use the task UVCONTSUB
\footnote{This step is completed prior to the release of
CASA version 5.6.1, and we use CASA version 5.4.0 when running UVCONTSUB.} 
to fit a linear function to the line-free channels, which models the continuum,
and subtract the continuum model to obtain the line-only visibility.
We then use the continuum data to perform phase self-calibration
and apply the self-calibration model to the line-only data.
We clean the continuum and line data with the CASA task TCLEAN 
using Briggs weighting, setting robust $=0.5$.
The synthesized beam has a size of $0\farcs31\times0\farcs27$ 
and a position angle of $39.4$ degrees.
Figure \ref{fig:clean} shows the cleaned image of the dust continuum 
and the zeroth, first, and second moments of the {\cii} emission.
J0439+1634 is clearly resolved as an arc-like shape,
which is typical for lensed galaxies.
The zeroth moment (integrated flux) of the {\cii} line
is more extended than the continuum flux.
The first moment (mean velocity map) shows ordered motion.

\begin{figure}
    \centering
    \includegraphics[width=0.99\linewidth]{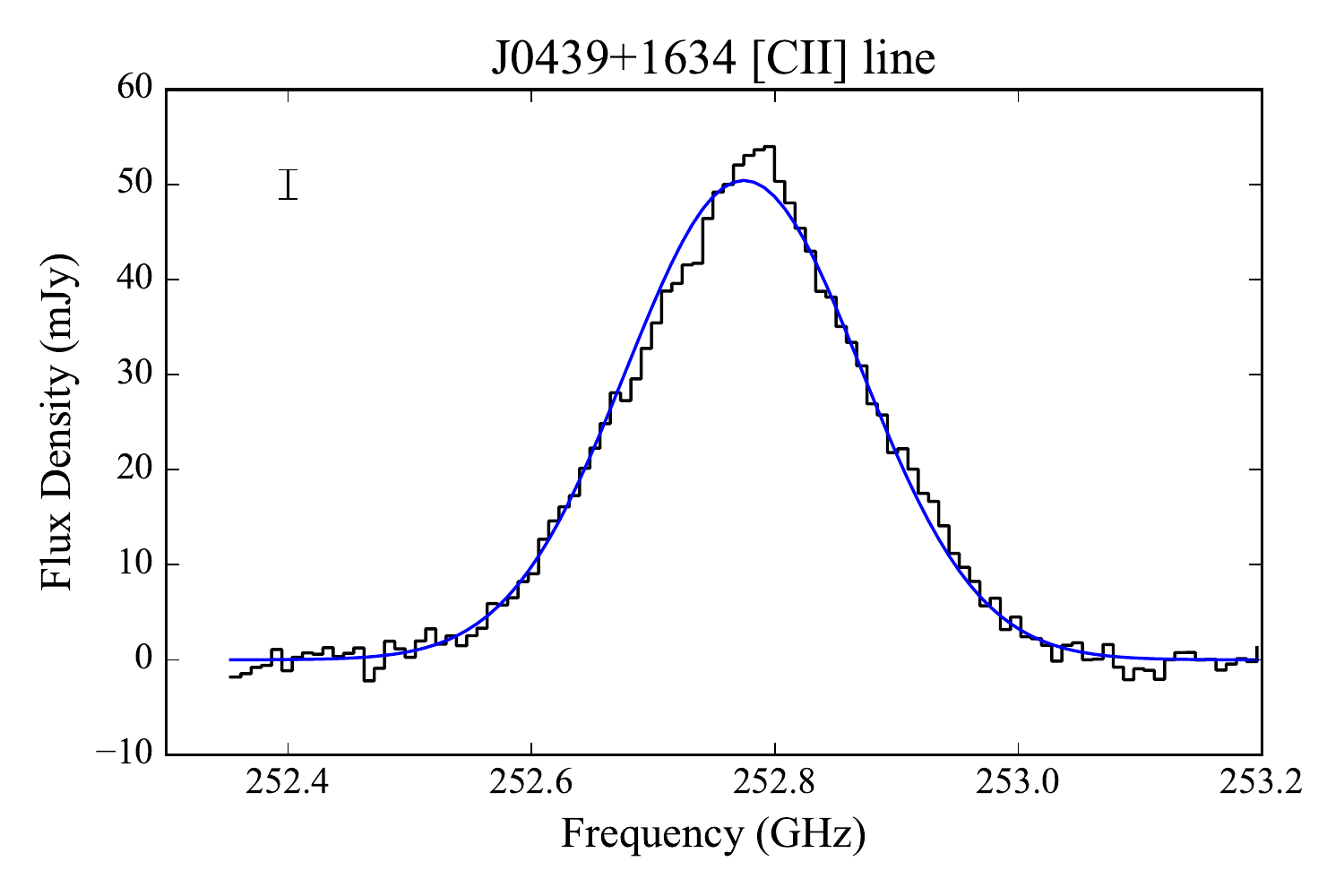}
    \caption{The {\cii} line profile of J0439+1634, 
    extracted from the cleaned image using a $2''$ diameter aperture. The error bar 
    in the upper left corner shows the uncertainty of the flux
    density. The red line presents the best-fit Gaussian profile,
    which has a FWHM of $270.0\pm2.8\text{ km s}^{-1}$ and a central frequency of $252.7744\pm0.0011$ GHz.}
    \label{fig:ciiprofile}
\end{figure}

We extract the continuum and {\cii} fluxes of J0439+1634 with a
$2\farcs0$ diameter aperture, which gives
$S_\text{245 GHz}=16.0\pm0.1 \text{ mJy}$ for the continuum
and $F_\text{\cii}=14.5\pm0.2\text{ Jy km s}^{-1}$ for the
integrated {\cii} flux.
We also fit a Gaussian profile to the {\cii} line using the CASA
task SPECFIT. The {\cii} line is centered at $252.7744\pm0.0011$ GHz
with an FWHM of $270.0\pm2.8 \text{ km s}^{-1}$,
which gives a redshift $z_\text{\cii}=6.51871\pm0.00003$.
In the rest of the paper, we set 252.7744 GHz as the rest frequency for {\cii}.
Figure \ref{fig:ciiprofile} shows the extracted {\cii} line profile,
which is well-fitted by a Gaussian function
and shows no evidence for an excess redshifted or blueshifted component.

\section{Lensing Model} \label{sec:lensing}

We use VISILENS \citep{spilker16}
to model the visibility of J0439+1634.
VISILENS is a parameterized lens modeling tool
for interferometry data. In short, VISILENS
models the $uv-$plane response of a lens system
and obtains the posterior distributions of model parameters
 using Markov Chain Monte Carlo (MCMC).

\begin{deluxetable*}{l|rrr|rrr}
\tablenum{1}
\tablecaption{{Lens model parameters}}
\label{tbl:lensparams}
\tablewidth{0pt}
\tablehead{
\colhead{} & \multicolumn{3}{c}{{Default}\tablenotemark{1}} & \multicolumn{3}{c}{ALMA only}\\
%\\\cline{2-4}  \cline{5-7}
\colhead{Parameters} & \colhead{lens} & \colhead{Continuum} & \colhead{{\cii}} & \colhead{lens} & \colhead{Continuum} & \colhead{{\cii}}
}
\startdata
Redshift & [0.67] & [6.5187] & [6.5187] & [0.67] & [6.5187] & [6.5187]\\
$\Delta \text{RA} ~('')$ \tablenotemark{2} & $0.459\pm0.002$ & $0.236\pm0.003$ & $0.235\pm0.002$ & $0.409\pm0.004$ & $0.222\pm0.005$ & $0.220\pm0.005$ \\
$\Delta \text{Dec} ~('')$ & $-0.050\pm0.001$ & $-0.019\pm0.001$ & $-0.017\pm0.001$ & $-0.058\pm0.002$ & $-0.019\pm0.003$ & $-0.017\pm0.003$\\
Mass $(M_\odot)$  \tablenotemark{3}& [$2.06\times10^{10}$] & - & - & $(1.95\pm0.03)\times10^{10}$ & - & - \\
$e$   \tablenotemark{4}& [0.65] & $0.409\pm0.011$ & $0.376\pm0.020$ & $0.641\pm0.007$ & $0.362\pm0.016$ & $0.347\pm0.021$\\
PA (deg) \tablenotemark{5}& [94.58] & $-6.9\pm1.2$ & $-0.6\pm1.7$ & $99.75\pm0.34$ & $4.9\pm1.5$ & $3.4\pm2.2$\\
Flux  \tablenotemark{6}& - & $3.46\pm0.04$ & $4.15\pm0.10$ & - & $3.18\pm0.07$ & $4.16\pm0.10$\\
$R_\text{eff} ~('')$  \tablenotemark{7}& - & $0.136\pm0.002$ & $0.233\pm0.006$ & - & $0.131\pm0.002$ & $0.235\pm0.007$ \\
$n$  \tablenotemark{8}& - & $1.71\pm0.06$ & $0.82\pm0.05$& - & $1.58\pm0.05$ & $0.91\pm0.05$\\
$\mu$  \tablenotemark{9}& - & $4.53\pm0.05$ & $3.44\pm 0.05$& - & $4.90\pm0.10$ & $3.45\pm 0.05$\\\hline
\enddata
\tablenotetext{1}{In the default model, the parameters of the lens galaxy,
except its position,  are fixed to the fiducial model in \citet{fan19}.}
\tablenotetext{2}{$\Delta \text{RA}$ and $\Delta \text{Dec}$
are relative to the phase center.}
\tablenotetext{3}{Mass of the lens galaxy.}
\tablenotetext{4}{Ellipticity of the lens or source.}
\tablenotetext{5}{Position angle (from north to east) of the lens or source. $\text{PA}=0$ means that the major axis lies east-to-west.}
\tablenotetext{6}{The source flux, in mJy for the continuum and in Jy km s$^{-1}$ for the {\cii} line.}
\tablenotetext{7}{The half-light radius.}
\tablenotetext{8}{The S\'ersic index.}
\tablenotetext{9}{The flux magnification.}
\tablecomments{The quantities in the square brackets are fixed. For the lens galaxy, 
\citet{fan19} use Einstein radius instead of mass. Here we follow the convention in VISILENS. 
The redshift and the mass of the lens galaxy are degenerate, 
and the lens parameters in this table gives the same lensing model as \citet{fan19}. 
Besides, note that the uncertainties only include statistical errors, and do not
take into account the systematic errors introduced by the model choice
(see Section \ref{subsec:systematics} for more discussion).}
\end{deluxetable*}

\subsection{Building the Lensing Model} \label{subsec:lensgalaxy}

%In their fiducial lensing model,
\citet{fan19} built the lensing model of J0439+1634 
based on the {\it HST} image, where
they used a singular isothermal ellipsoid \citep[SIE; e.g.,][]{sie}
to describe the mass distribution of the lens galaxy.
In their fiducial model, 
the lensing galaxy has a high ellipticity $(e=0.65)$,
lies at the west side of the quasar,
and generates three quasar images.
The position, the ellipticity, and the position angle 
of the modeled lens galaxy are consistent with the observed {\it HST} optical image.

The {\it HST} images have spatial resolution of $\sim 0\farcs075$,
which is several times better than the current ALMA data.
We thus adopt the lens mass distribution 
from the fiducial model in \citet{fan19}.
Specifically, we use an SIE to describe the lens galaxy. 
The Einstein radius, 
ellipticity, and position angle of the SIE are fixed 
to the values in the fiducial {\it HST} model, while
the position of the lens is left free, which
accounts for any pointing offsets between {\it HST} and ALMA.
Because the continuum has a higher signal-to-noise ratio (SNR),
we first fit the continuum to obtain the best-fit lens position,
then apply the lens position when fitting the {\cii} emission.
We use a S\'ersic profile to describe the source,
 both for the continuum and the {\cii} line emission.
This model is referred to as the default model in this paper.

For comparison, we also build an alternative model,
hereafter referred as the ``ALMA-only" model, in which  we leave all parameters free
when fitting the ALMA data and do not use any information from the {\it HST} observations. 
Again, we use an SIE to describe the lens galaxy and a S\'ersic profile
to describe the source emission, both for the continuum and the {\cii} line.
We first fit the continuum to obtain the best-fit values
of the lens parameters,
then apply these values when fitting the {\cii} line.

\begin{figure*}
    \centering
    \includegraphics[width=0.9\textwidth]{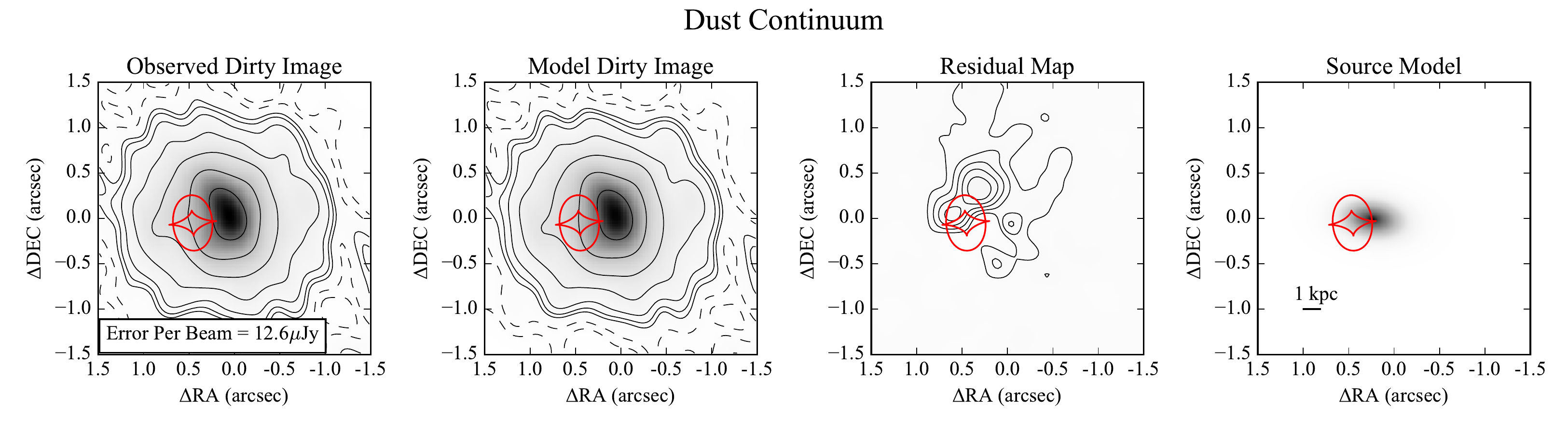}
    \includegraphics[width=0.9\textwidth]{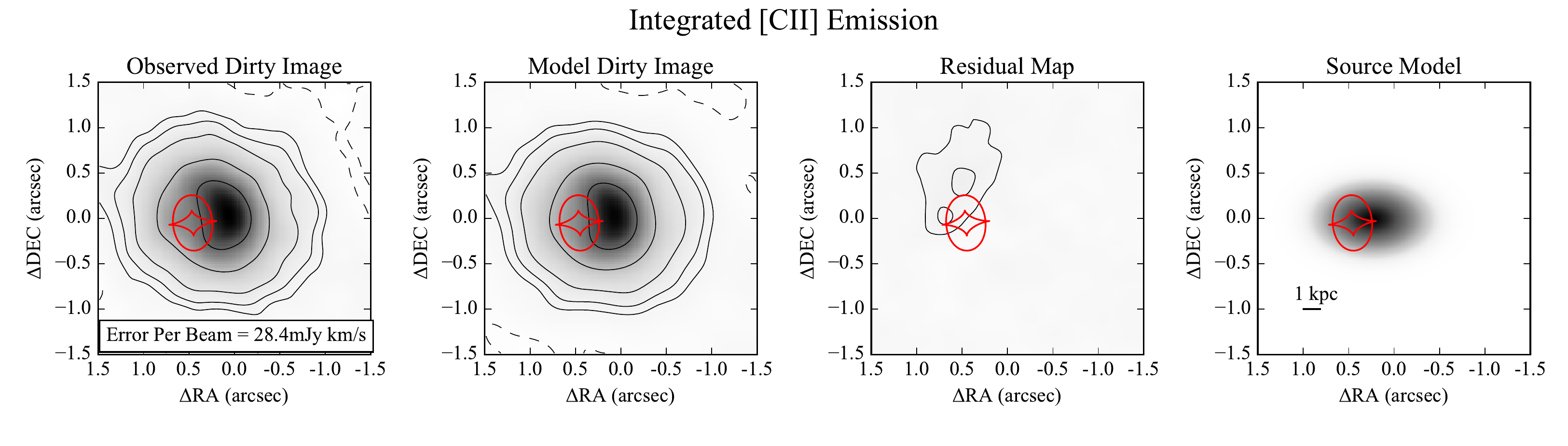}
    \caption{
    The default lensing model of J0439+1634 based on ALMA and {\it HST} observations.
    {\it Upper Panel}:
    the continuum model. From left to right: the observed dirty image with natural weighting, 
    the modeled dirty image with natural weighting,
    the residual image, and the source model.
    Contours in the observed and model images 
    are $-10\sigma$, $-5\sigma$, $5\sigma$, $10\sigma$, $20\sigma$,
    $50\sigma$, $100\sigma$, $200\sigma$ and $400\sigma$ levels,
    where $1\sigma$ equals to the ``error per beam" 
    in the observed image.
    Contours in the residual image
    are $-5\sigma$, $5\sigma$, $10\sigma$, $15\sigma$, $20\sigma$ and
    $25\sigma$ levels. 
    In all images, dashed black lines are negative contours and solid black lines are positive ones.
    The red line marks the caustics of the lens.
    When fitting the continuum, we fix the lens parameters, except the position,
    to the fiducial model in \citet{fan19}.
    {\it Lower Panel}: Same as the upper panel, but for {\cii} emission.
    The lens parameters in the {\cii} fitting are fixed to the best-fit values in the continuum model.}
    \label{fig:lensmodel}
\end{figure*}

\begin{figure*}
    \centering
    \includegraphics[width=0.9\textwidth]{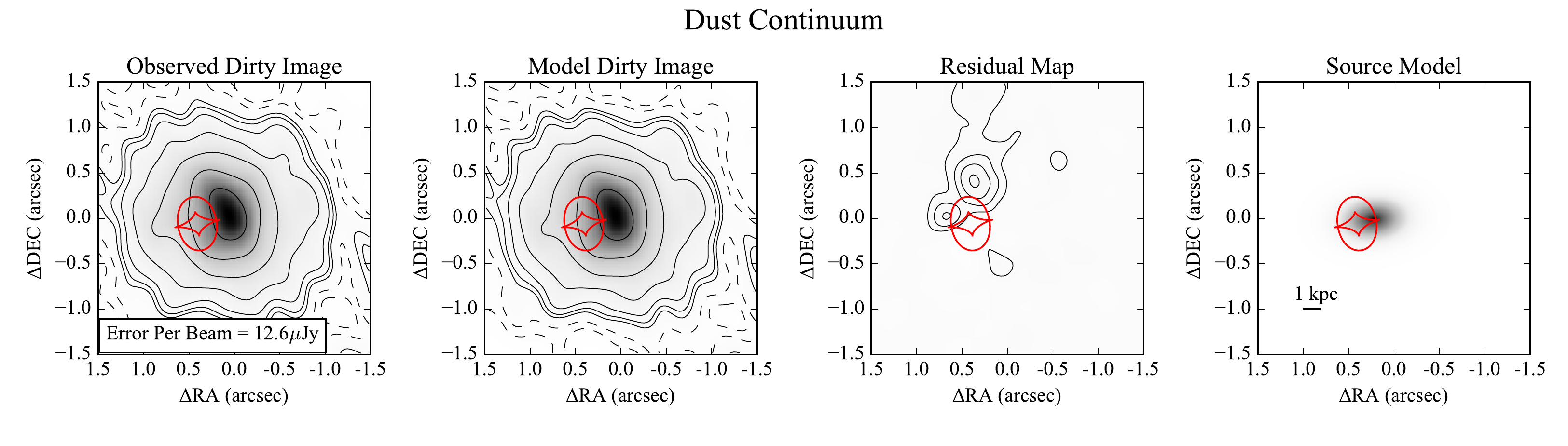}
    \includegraphics[width=0.9\textwidth]{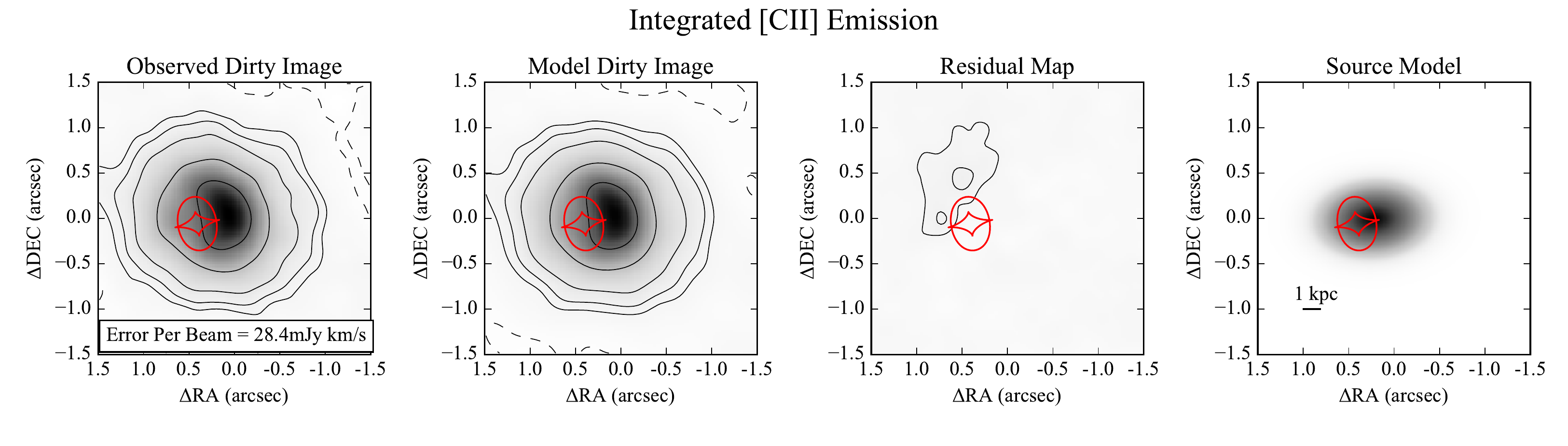}
    \caption{Same as Figure \ref{fig:lensmodel},
    but for the ALMA-only lensing model.
    We first fit the continuum visibility with all parameters left free,
    then fit the {\cii} emission with lens galaxy parameters fixed to the best-fit values in the continuum model.
    This ALMA-only model and the default model in Figure \ref{fig:lensmodel}
    are nearly identical.
    }
    
    \label{fig:testmodel}
\end{figure*}

Figure \ref{fig:lensmodel} shows the fitting result of the 
default model, and Figure \ref{fig:testmodel} illustrates the 
ALMA-only model.
Table \ref{tbl:lensparams} summarizes the best-fit parameters
for both models.
%Here we illustrate dirty images, because dirty images are the most equivalent comparison from the visibility-based fitting.
Despite tiny differences in details, 
the two models give the same overall lensing structure.
%We mainly care about the flux and the size of the quasar host galaxy.
%For these quantities, the two models are consistent within $2\sigma$.
Because the {\it HST} images have better resolution,
we use the default model to derive the properties of J0439+1634 from this point on.
We will discuss the systematic errors introduced by the choice
of the model in Section \ref{subsec:systematics}.

In addition to the fiducial model, \citet{fan19} raise
two alternative models, in which the lens mass distribution
differs significantly from the fiducial model and produces either double or quadruple quasar images.
See Figure 4 in \citet{fan19} for more information.
The alternative models do not provide suitable fits to the ALMA data. 
We conclude that the fiducial {\it HST} model has the correct lensing configuration.

\subsection{Dust Continuum} \label{subsec:cont}

The upper panel of Figure \ref{fig:lensmodel} shows the best-fit
dust continuum in the default model.
As described in Section \ref{subsec:lensgalaxy}, 
in the default model, we fix the deflector galaxy mass distribution to the fiducial model
in \citet{fan19} which is based on {\em HST} imaging, and fit the quasar host galaxy emission
in ALMA data as a S\'ersic profile.
The dirty images are generated with natural weighting
to enhance the SNR.
The dust continuum of J0439+1634 can be well-fitted by a single S\'ersic
profile, with a reduced $\chi^2=1.035$. 
The best-fit S\'ersic index is $1.71\pm0.06$ 
and the half-light radius is $0\farcs136\pm0\farcs002$ 
($0.74\pm0.01$ kpc),
suggesting a compact, exponential-disk-like profile.
(See Section \ref{subsec:contcii} for further discussion.)
The overall magnification is $4.53\pm0.05$ when averaged over the entire galaxy. 
Compared to the fiducial {\it HST} model, 
the position of the optical quasar
deviates from the continuum center by $0\farcs014$.
The typical astrometric error for ALMA is about 5\% of the resolution, 
which translates to $\sim0\farcs015$ given a beam size of $\sim 0\farcs3$
\citep[the ALMA technical handbook, e.g., ][]{almatech}.
The positions of the optical quasar and the host galaxy are thus consistent.

%{\bf add a couple of sentence describing what the best fit model means: size of the galaxy - compact, exponential profile etc., and forward reference to \S 4.1}

%{\bf I suggest that we add the magnification map to Figure 5 as the fourth panel, and devote a paragraph to discuss the differential magnification, and give average magnifications with in the central 200, 500 and 1000 pc here (in both arcsec and pc), to demonstrate that we are resolving the structure close to the nucleus of the galaxy. Also, make the black cross more prominent in the figure.}

The residual map shows some statistically significant structures.
The peak of these structures is $4.9\%$ of the peak in 
the observed dirty image.
We expect such features given that 
we use a simple SIE + S\'ersic model
and the SNR of the data is high (with natural weighting,
the peak SNR in the dirty image is $\sim600$).
When we add a Gaussian profile to the source model,
where we allow the Gaussian profile to have negative flux,
 the flux of the Gaussian profile converges to zero within the error. 
We thus argue that the structures in the residual image
cannot be explained by a single bump or void in the source galaxy.
The structures might result from an over-simplification
of the lens and source model.

\subsection{Integrated {\cii} Flux} \label{subsec:CIIsersic}

The lower panel of Figure \ref{fig:lensmodel} shows the best-fit 
result of the {\cii} emission in the default model,
and Table \ref{tbl:lensparams} shows the parameters of {\cii} line observations.
The reduced $\chi^2$ of the best-fit model is $1.014$.
%{\bf add a sentence describing something like the size, profile, inclination of the CII emission.}
The best-fit {\cii} emission has a S\'ersic index of $0.82\pm0.05$, 
consistent with an exponential $(n=1)$ profile, 
and a half-light radius of $0\farcs233\pm 0\farcs006$ $(1.27\pm0.03 \text{ kpc})$.
The position and the ellipticity 
of the integrated {\cii} emission are consistent 
with those of the dust continuum within $2\sigma$, 
while the {\cii} line has a smaller S\'ersic index 
and larger half-light radius.
This difference suggests that the {\cii} line is more diffuse than the dust,
as shown in the clean images.
The overall magnification of the {\cii} emission is $3.44\pm0.05$,
which is smaller than that of the dust continuum,
mainly because the {\cii} profile is more diffused.

Similar to the continuum, a S\'ersic profile captures the major features of 
the {\cii} emission.
The residual image of the {\cii} emission 
is similar to the continuum residual.
The peak in the residual is $8.1\%$ of the peak in the dirty image.
%The nature of these features might be complicated and
%we refer it to future studies.

\subsection{{\cii} Kinematics} \label{subsec:kinematics}

Figure \ref{fig:clean} suggests that the host galaxy of J0439+1634 
has an ordered, rotation-like velocity field. We thus fit the {\cii} emission
using an axisymmetric rotating thin disk,
following the method described in \citet{neeleman19}.
In short, we set up parameterized models for the flux distribution,
the mean velocity field, and the velocity dispersion field.
We then use VISILENS to calculate the lensed {\cii} emission 
and the $uv-$plane response in each velocity channel.
We obtain the best-fit model parameters by minimizing 
the residual of the visibility in all channels.
To keep maximum flexibility, we do not constrain the parameters using 
the S\'ersic model for the integrated {\cii} flux.
We assume a S\'ersic profile for the flux distribution
and apply various forms for the rotation curve and the velocity dispersion profile.
However, all of these models return large residuals and unphysical best-fit parameters.
We thus conclude that J0439+1634 cannot be described by an axisymmetric
rotating thin disk.

The main reason for the poor fit is the apparent misalignment between
the major axis of the flux distribution and the velocity gradient.
For an axisymmetric rotation disk, the major axis and the velocity gradient 
should be in the same direction. In contrast, the major axis of 
the flux distribution of J0439+1634
is roughly aligned east-to-west (Figure \ref{fig:lensmodel},
right panel), 
while the velocity gradient is roughly north-to-south (Figure \ref{fig:clean}).
To further investigate this problem, we estimate the source-plane flux distribution
using a simple inverse ray-tracing method. 
Specifically,
we reconstruct the source (i.e., un-lensed) data cube on a grid
with a pixel size of $0\farcs04$.
Using the overall {\cii} magnification $\mu_\text{\cii}=3.44$, 
we estimate the average source-plane resolution to be 
$\sim\sqrt{0\farcs31\times0\farcs27}/\sqrt{3.44}=0\farcs156$ 
for the {\cii} emission.
A pixel size of $0\farcs04$ gives a super-Nyquist sampling,
which helps to resolve the regions with higher magnification
than the average value.
We then trace all the pixels in the image-plane data cube (i.e., the clean image)
to the source plane according to the default lensing model.
If more than one image-plane pixels are traced to
the same source-plane pixel, these image pixels are averaged.
This simple method captures the main features of the quasar host galaxy
without expensive pixelized lensing reconstruction.
%We also calculate the error of the reconstructed source data cube 
%based on the error of the image data cube.

%These features suggest that J0439+1634 has a rotation-like velocity field, 
%but is more complicated than an axisymmetric rotating thin disk. 
%Many quasar host galaxies at $z\gtrsim6$ show similar features \citep[e.g.,][]{pensabene20}.

\begin{figure*}
    \centering
    \includegraphics[width=1\textwidth,trim=0cm 0cm 0cm 0cm,clip]{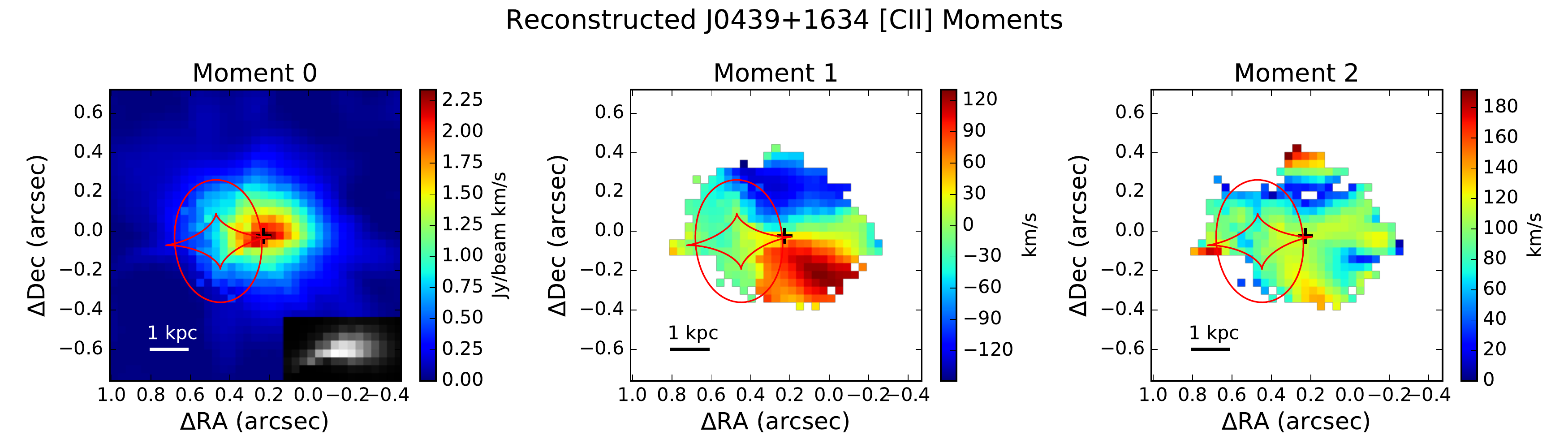}
    \caption{The reconstructed source-plane moments. 
    From left to right: the reconstructed zeroth, first, and second moments.
    The coordinates are relative to the phase center.
    The images have pixel sizes of $0\farcs04$ (0.22 kpc).
    The black cross marks the center of 
    the {\cii} emission from the default lensing model.
    The red line illustrates the lensing caustics. The moment 2 map has some pixels
    missing due to low SNR which leads to mathematical errors 
    (i.e., getting a squared root of a negative value).
    {The lower-right corner of the left panel shows the output
    when performing the inverse ray-tracing analysis to a beam on the image plane 
    which locates at the flux peak. The ``reconstructed" beam has a size of
    $0\farcs26\times0\farcs09$ and is a rough estimate of the source-plane beam shape.}
    }
    \label{fig:reconstruct}
\end{figure*}

We generate source-plane moment maps using the reconstructed data cube.
When calculating the first and the second moments,
we only include pixels that have integrated flux SNR larger than $10$.
%We also estimate the error of the moments according to the noise cube.
Figure \ref{fig:reconstruct} shows the reconstructed moments,
which confirm the overall picture of J0439+1634:
a regular profile for the integrated emission (moment 0)
and a rotation-like mean velocity field (moment 1).
The major axis of moment 0 is significantly offset from the 
velocity gradient in the moment 1 map, confirming the argument 
we made with the lensed image.
Another hint is the structures in the moment 2 map. 
 For a rotating thin disk,
we expect a peak at the center of the moment 2 map due to the beam smearing effect
where the line-of-sight velocity gradient is large.
This peak is not seen in Figure \ref{fig:reconstruct}; instead, the moment 2 map
has complex structures, which indicate complicated velocity field in the host galaxy.

The lower-right corner of the reconstructed moment 0 map illustrates the output 
when we perform the inverse ray-tracing analysis to an image-plane beam 
located at the image-plane flux peak. This ``reconstructed" beam 
is a rough estimate of the beam shape on the source plane. 
%Lensing squeezes the source-plane beam in the north-south direction and 
%introduces significant distortions. 
The source-plane beam has a size of $0\farcs26\times0\farcs09$,
which further illustrates that a source-plane pixel size of $0\farcs04$ is appropriate.

\begin{figure*}
    \centering
    \includegraphics[width=0.8\textwidth,trim=0cm 0cm 0cm 0cm,clip]{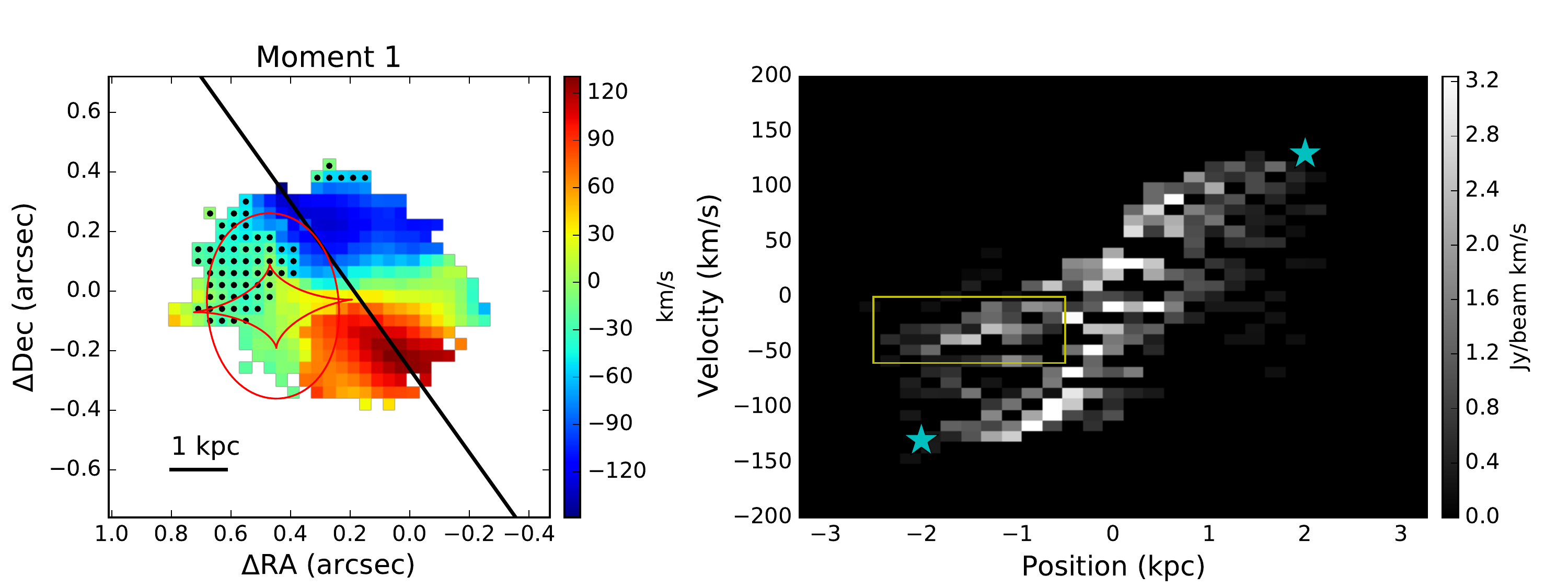}
    \caption{{\it Left:} The moment 1 map. The black line illustrates the axis used to
    extract the position-velocity diagram. The axis connects the pixels with 
    the maximum and minimum moment 1 value (i.e., the blue peak and the red peak).
    The black dots mark the pixels that contributes to the minor feature in the 
    position-velocity diagram (marked by the yellow box in the right panel). The red curve
    marks the caustics. {\it Right:} The position-velocity diagram. The diagram contains
    a rotation-like feature and a minor feature marked by the yellow box.
    The two cyan asterisks correspond to the value we adopt for the 
    maximum velocity and the corresponding radius:
    $v_\text{max}\sin(i)=130 \text{ km s}^{-1}$ and $r_\text{max}=2\text{ kpc}$.}
    \label{fig:pvplot}
\end{figure*}

Note that the blue and red wings in the moment 1 map 
are located outside of the caustics, which means that they are not multiply-imaged. 
For these areas,
the effect of gravitational lensing on the observations is equivalent to
shrinking the beam size and applying some image distortions.
As such, the inverse ray-tracing reconstruction
can capture the structure of the velocity field,
especially in the blue and red wings.
In Figure \ref{fig:pvplot}, we illustrate the position-velocity plot of J0439+1634,
generated using the reconstructed moment 0 and moment 1.
We extract the velocities along the black line, which connects the pixels with 
the maximum and minimum moment 1 value (i.e., the red and blue peaks).
The position-velocity plot clearly shows a rotation-like feature.
The velocity rises at $r\lesssim1\text{ kpc}$ and flattens beyond this radius.
The maximum rotation velocity is roughly $v_\text{max}\sin(i)=130\text{ km s}^{-1}$
where $i$ is the inclination angle of the rotation axis, and the velocity is
measured out to $r_\text{max}=2\text{ kpc}$.
%These two values are marked by the cyan asteroids in the position-velocity plot.

In addition to the rotation-like feature,
the position-velocity plot contains another component with low velocity 
$(|v|\lesssim50\text{ km s}^{-1}, \text{marked by the yellow box})$.
We mark pixels that contribute to this feature with black dots 
in the moment 1 map in Figure \ref{fig:pvplot}.
These pixels either lie inside the caustics,
which means that they are multiply imaged, and have complicated lens-mapping function,
or have low signal and large errors.
The simple inverse ray-tracing method will fail for these areas, 
and this component is likely an artifact.
If it is physical, it might reflect some complex structures in the host galaxy,
which can be resolved in the upcoming high-resolution ALMA observations.
Possible scenarios include minor mergers and clumpy star-formation regions.
In any case, this structure only contributes $\sim10\%$ of the total flux,
and the rotation-like feature dominates the velocity field.

J0439+1634 has a rotation-like velocity field, 
but cannot be described by an axisymmetric rotating thin disk.
It is likely that the {\cii} emission is not axisymmetric and/or
that the host galaxy has a thick geometry.
Both cases are common for high-redshift sgalaxies
\citep[e.g., ][]{pensabene20, fsw20}.
Specifically, many star-forming galaxies at $z\gtrsim2$
have velocity dispersion $\sigma\gtrsim45 \text{ km s}^{-1}$ and show thick geometry
\citep[for a recent review, see][]{fsw20}.
For J0439+1634, we can estimate its velocity dispersion using the regions 
where the rotation curve has flattened. %Here we use the pixels that have 
The moment 2 map suggests that these regions have 
$\sigma\sim70 \text{ km s}^{-1}$, which means that a thick geometry is likely.

Fitting a non-axisymmetric model and/or a thick disk model requires high 
spatial resolution, and our data cannot put a strong constraint on these models.
In future work, we will perform pixelized lensing reconstruction
and detailed dynamical modeling once the high-resolution ALMA observations
are carried out (Project 2018.1.00566.S, PI: Fan).

%As suggested by the reconstructed source-plane moment 1 map,
%the maximum line-of-sight velocity is $149\text{ km s}^{-1}$,
%consistent with the values in the image-plane (i.e., lensed) moment 1
%map shown in Figure \ref{fig:clean}.
%The distance of the corresponding pixels to the center of the galaxy is .... 
%Given the complexity of the {\cii} emission, 

%\citet{dye15} argue that reconstructing the source
%using the clean image gives consistent results with working
%on the $uv-$plane, and this method has been adopted in recent studies
%\citep[e.g.,][]{cheng20}.

\subsection{Systematic Uncertainties} \label{subsec:systematics}

We first consider the systematic errors in the fluxes and sizes of
the continuum and the {\cii} emission.
The main source of systematic errors 
is that the SIE + S\'ersic lensing model is over-simplified.
As a result, there are positive clumps
in the residual images in Figure \ref{fig:lensmodel}.
The flux of the clumps is much smaller than the
the flux calibration error  
and is negligible in the error analysis.
Properly modeling the structures in the residual images
requires expensive pixelized modeling of the source flux and 
the lens mass \citep[e.g.,][]{hezaveh16},
which is beyond the scope of this paper.

We then consider the systematic uncertainties in the source-plane reconstruction
in Section \ref{subsec:kinematics}.
The major uncertainty is the beam-smearing effect.
The source-plane resolution is $\sim0\farcs15$ $(\sim0.8\text{ kpc})$.
In this study, we focus on the maximum rotation velocity
rather than detailed velocity field structure.
Beam-smearing effects have little influence on our main result, 
because (1) the blue and red peaks are only singly imaged, and
(2) the rotation velocity flattens at $r\gtrsim 1 \text{ kpc}$,
so the central, low-velocity area does not influence the edge at $r_\text{max}=2\text{ kpc}$
where we measure the maximum velocity.
Similar methods have been adopted by recent studies
to measure the rotation velocity of lensed galaxies \citep[e.g.,][]{cheng20}.

%%%%%%%%%%%%%%%%%%%%%%%%%%%%%%%%%%%%
%%%%%%%%%%%%%%%%%%%%%%%%%%%%%%%%%%%%
%  Section 4: Physical Properties  %
%%%%%%%%%%%%%%%%%%%%%%%%%%%%%%%%%%%%
%%%%%%%%%%%%%%%%%%%%%%%%%%%%%%%%%%%%

\section{Physical Properties of J0439+1634} \label{sec:physical}

\subsection{Dust Continuum and {\cii} Emission} \label{subsec:contcii}

The host galaxy of J0439+1634 has a regular S\'ersic profile,
both for the continuum and {\cii} emission.
The S\'ersic index is close to one,
which suggests that J0439+1634 is more similar to an exponential disk
than a de-Vaucouleurs bulge with $n=4$.
The {\cii} line profile is also well-described by a single-peaked
Gaussian profile, with no excess of blueshifted or redshifted components.
The smooth structures in the moment maps and the position-velocity plot 
disfavor the scenario of a close, on-going major merger.
In addition, no other objects are detected within the ALMA field of view 
$(\sim12'')$.
These results suggest that J0439+1634 is not an
on-going major merger and does not exhibit significant outflow features in the {\cii} velocity field. 
However, it is possible that J0439+1634 is
a minor merger or a remnant of a recent major merger.

%There is no strong evidence for outflows and/or signs of merger.

%We first compare our measurement of the continuum and {\cii}
%flux with the literature.
%Based on the James Clerk Maxwell Telescope (JCMT)
%and the NOrthern Extended Millimeter Array (NOEMA) observations,

\citet{yang19} measure the far-infrared (FIR) to centimeter-wavelength
spectral energy distribution (SED), as well as the
CO, {\ci}, {\cii}, {\oi}, and $\text{H}_2\text{O}$ emission lines 
of J0439+1634. 
Their NOrthern Extended Millimeter Array (NOEMA)
observation gives $z_\text{\cii}=6.5188\pm0.0002$,
$F_\text{\cii}=11.7\pm0.8\text{ Jy km s}^{-1}$,
$\text{FWHM}_\text{\cii} = 328.1\pm18.0 \text{ km s}^{-1}$,
and $S_\text{239 GHz}=14.0\pm0.1 \text{ mJy}$.
The typical flux calibration uncertainty is $\sim 15\%$ for NOEMA
and $\sim 10\%$ for ALMA.
The continuum and {\cii} line fluxes reported
in Section \ref{sec:data} are consistent with those in \citet{yang19}.
The difference in the FWHM of the {\cii} line is about 3$\sigma$.

Based on these measurements,
\citet{yang19} calculate the infrared luminosities,
emission line luminosities, star formation rate, dust mass,
and gas mass without correcting for the lensing magnification.
We refer the reader to \citet{yang19} for the details 
of how these properties are calculated.
In this work, with spatially resolved ALMA images, 
we calculate the de-lensed values of these quantities.
For {\cii}-based quantities, we apply the magnification of {\cii} emission,
$\mu_\text{\cii}=3.44$;
for FIR-based quantities, we apply the continuum magnification, 
$\mu_\text{cont}=4.53$.
In addition, we apply the continuum magnification to 
the CO-based molecular gas mass, assuming that 
the dust continuum traces the molecular gas.
The results are listed in Table \ref{tbl:physical}.
Similar to other high-redshift quasars \citep[e.g.,][]{walter09,decarli18,wang19b},
J0439+1634 is hosted by a gas-rich ultra luminous infrared galaxy (ULIRG)
with intense star formation activity,
with a TIR luminosity of $\sim10^{13}L_\odot$.

\begin{deluxetable}{lr}
\tablenum{2}
\label{tbl:physical}
\tablecaption{De-lensed Physical Properties of J0439+1634}
\tablewidth{0pt}
\tablehead{
\colhead{{\cii}-Based Properties} & \colhead{}}
%\decimalcolnumbers
\startdata
$L_\text{\cii} (L_\odot)$ &  $(3.5\pm0.3)\times10^9$ \\
$L'_\text{\cii} (\text{K km s}^{-1}\text{pc}^{2})$ & \phantom{123456789012345} $(1.7\pm0.1)\times10^{10}$ \\
$M_{\text{C}^+}(M_\odot)$ & ($1.1\pm0.1)\times10^7$\\
$\text{SFR}_\text{{\cii}} (M_\odot\text{ year}^{-1})$ & $\sim233\sim1398$\\\hline
\hline
Other properties & \\
\hline
$L_\text{FIR} (L_\odot)$& $(7.5\pm0.4)\times10^{12}$\\
$L_\text{TIR} (L_\odot)$& $(1.06\pm0.04)\times10^{13}$\\
$\text{SFR}_\text{TIR} (M_\odot\text{ year}^{-1})$ & $1.56\times10^3$\\
$M_\text{dust}(M_\odot)$ & $4.9\pm0.2\times10^8$\\
$M_{\text{H}_2,\text{CO}}(M_\odot)$ & $1.19\times10^{10}$\\\hline
\enddata
\tablecomments{
These values are calculated according to Table 1 in \citet{yang19}.
We apply $\mu=\mu(\text{\cii})=3.44$ for {\cii}-based quantities and
$\mu=\mu(\text{continuum})=4.53$ for the other quantities.
FIR luminosity includes flux in rest-frame $42.5-122.5~\mu$m,
and TIR luminosity includes flux in rest-frame $8-1000~\mu$m.
The uncertainties only reflect statistical errors.
}
\end{deluxetable}

One interesting aspect of the {\cii} emission is the so-called ``{\cii} deficit."
In systems with high FIR surface brightness
\citep[$\Sigma_\text{FIR}\gtrsim10^{11}L_\odot \text{ kpc}^{-2}$; e.g.,][]{diaz-santos17, hc18},
the {\cii}-to-FIR ratio decreases with $\Sigma_\text{FIR}$.
This relation has been observed in many different systems,
including nebulae in the Milky Way 
\citep[for example, the Orion Nebula; e.g.,][]{goicoechea15},
local luminous infrared galaxies (LIRGs) and ULIRGs
\citep[e.g.,][]{diaz-santos13},
high-redshift submillimeter galaxies (SMGs)
\citep[e.g.,][]{oteo16, spilker16, litke19}
and quasars \citep[e.g.,][]{decarli18,neeleman19}.
The underlying mechanism of the {\cii} deficit might be complex.
Some plausible scenarios include:
(1) In regions with higher surface density, 
most carbon atoms are in the form of CO molecules rather than 
 $\text{C}^+$ ions \citep{narayanan17};
(2) {\cii} might become optically thick at high surface density \citep{luhman98};
(3) Large FIR surface brightness indicates strong dust absorption
of the UV radiation, which is the main heating source of {\cii} emission 
\citep{hc18};
(4) {\cii} emission might be saturated in warm gas
\citep{mo16}.

%{\cii} lines in high-redshift quasars comes predominantly
%from the warm photon-dissociated regions (PDRs)
%which are heated by UV radiation generated by star formation
%or AGN activity.

%{\bf refer to magnification map to justify that we are resolving the distribution.}
With spatially resolved data, %
we can investigate the {\cii} deficit in different regions of J0439+1634.
Using the continuum and {\cii} model in Figure \ref{fig:lensmodel},
we measure the {\cii} and FIR flux of J0439+1634 
in four regions.
The $n^\text{th}$ region is defined as
$n\times R_\text{eff}<R<(n+1)\times R_\text{eff}$,
where $n=0,1,2,3$, and $R_\text{eff}=0\farcs136$ 
is the half-light radius of the continuum emission. 
All regions have the same center, ellipticity, and position angle as the 
continuum emission. 
The widths of the rings are close to the average source-plane
resolution of $0\farcs15$.
%The physical widths of each ring is about $0\farcs07$ or $\sim350\text{ pc}$,
%close to the source plane resolution.
The flux calibration error ($\sim15\%$)
is the dominant source of uncertainty, and we ignore other uncertainties.

%{\cii} emission lines can be generated under various ISM conditions.

%All three annuli falls on the {\cii}-FIR deficit 
%with decreasing $\text{\cii}/\text{FIR}$ towards large radii.
%This trend explains that the {\cii} emission is more diffused than
%the continuum. 

\begin{figure}
    \centering
    \includegraphics[width=0.48\textwidth]{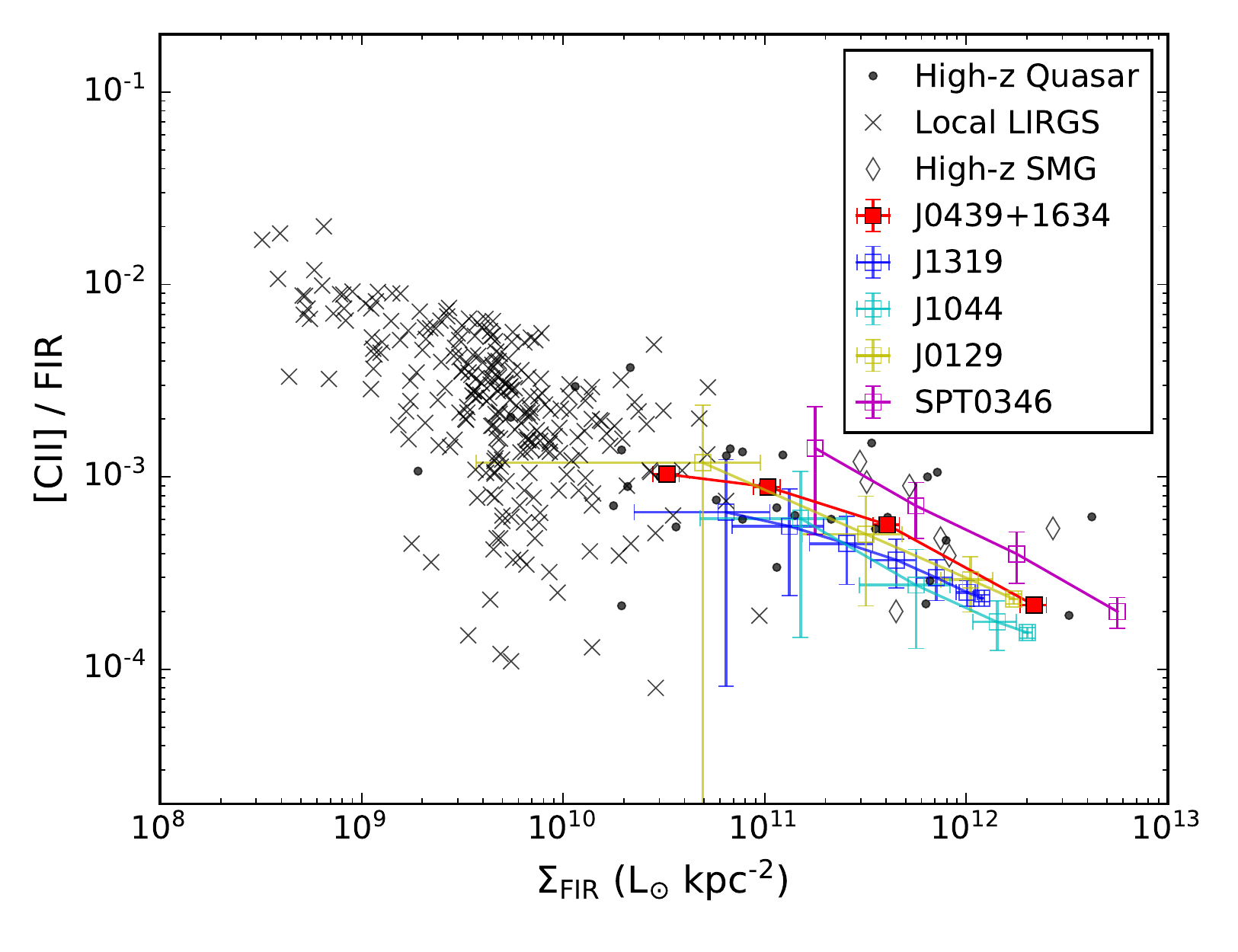}
    \caption{The {\cii}-FIR ratio versus FIR surface brightness plot
    for resolved regions in J0439+1634, local LIRGs in the GOALS
    sample \citep{diaz-santos13},
    quasars at $z\gtrsim6$ 
    \citep{wangran13, venemans16, venemans17, shao17, decarli18, izumi18, wang19b, wangran19, venemans20}, and
    $4.5<z<6.5$ SMGs \citep{riechers13, neri14, gullberg18, litke19}.
    Besides J0439+1634,
    three quasars and an SMG in this plot have resolved measurements,
    including J1319 \citep{shao17}, 
    J0129, and J1044 \citep{wangran19}, and SPT0346 \citep{litke19}.
    The points representing different
    regions in one object are connected by a line.}
    \label{fig:ciifir}
\end{figure}

Figure \ref{fig:ciifir} illustrates the position of J0439+1634
on the {\cii}$/$FIR$-\Sigma_\text{FIR}$ plot.
We include $z\gtrsim6$ quasars from the literature for comparison.
Specifically, three quasars from \citet{shao17} and \citet{wangran19} have
spatially resolved measurements. 
%For these quasars and J0439+1634,
%we connect the points that represent different regions by lines.
We also include
LIRGs from the Great Observatories All-sky LIRG Survey (GOALS) sample 
\citep{diaz-santos13}, SMGs at $4.5<z<6.5$
\citep{riechers13,neri14,gullberg18},
and resolved regions of a gravitationally lensed SMG at $z=5.7$, SPT0346
\citep{litke19}.
Similar to other objects with high-resolution data, 
 different regions in J0439+1634 tightly follow the {\cii} deficit.
Our result thus suggests that the {\cii} deficit is
related to physical processes on $\lesssim1 \text{ kpc}$ scales,
i.e., it reflects the  properties of the local ISM 
rather than the entire galaxy.

\subsection{Host Galaxy Dynamics} \label{subsec:dynamics}

Using the maximum line-of-sight velocity $v_\text{max}\sin(i) = 130 \text{ km s}^{-1}$
and the corresponding radius $r_\text{max}=2\text{ kpc}$,
we estimate the dynamical mass within $r_\text{max}$ for J0439+1634,
$M_\text{dyn}\sin^2(i)=7.9\times10^{9}M_\odot$.
We report in Table \ref{tbl:physical} the $\text{H}_2$ mass of J0439+1634,
$M_{\text{H}_2,\text{CO}}(M_\odot)=1.19\times10^{10} M_\odot$,
which gives a gas-mass fraction of $1.5\times\sin^2(i)$.
The gas-mass fraction is high except for very low inclinations.

Although the host galaxy of J0439+1634 has a regular shape and a rotation-like velocity field,
we show in Section \ref{subsec:kinematics} that this galaxy is not an axisymmetric
thin disk. Plausible scenarios include (1) the emission in the galaxy is not 
axisymmetric, which can happen when the star-forming regions are not
evenly distributed on the disk, or (2)
the host galaxy has a ``thick" geometry (e.g., a spheroid) and is not a thin disk.
The upcoming high-resolution ALMA data will reveal small-scale structures
that might distinguish these models.
Here we briefly discuss the implications of the thick disk model, since
the moment 2 map indicates that a thick geometry is likely (Section \ref{subsec:kinematics}).

A thick disk with a large velocity dispersion has
a non-negligible turbulent pressure gradient which needs to be considered when estimating
the dynamical mass. Following the discussion in \citet{fsw20},
we estimate the circular velocity of the host galaxy, $v_c$:
\begin{equation}
    v_c^2=v_\text{max}^2+2\sigma^2\times(r/R_d)
\end{equation}
which is related to the dynamical mass by $M_\text{dyn}(r)=v_c^2r/G$,
where $R_d$ is the scaling radius of the exponential disk (i.e., $I(r)\propto e^{-r/R_d}$).
Applying $v_\text{max} = 130/\sin(i) \text{ km s}^{-1}$, $r=2 \text{ kpc}$, 
$\sigma=70 \text{ km s}^{-1}$ and $R_d=0.8$ kpc according to the best-fit {\cii} model
in Section \ref{subsec:CIIsersic} yields
\begin{equation}
    v_c^2=\left[\frac{130 \text{ km/s}}{\sin(i)}\right]^2 + (160\text{ km/s})^2
\end{equation}
which suggests that, in the thick-disk model,
the contribution of velocity dispersion is significant.
However, the velocity dispersion should be taken as an upper limit
given the beam smearing effect.
It is hard to correct the beam smearing effect under the current resolution; as such,
we still use $M_\text{dyn}\sin^2(i)=7.9\times10^{9}M_\odot$ in the rest of this paper.
This result illustrates the need of careful modeling with high-resolution data
when measuring the dynamical mass of high-redshift quasar host galaxies.

\citet{pensabene20} analyzes the archival ALMA data of 32 quasars to model their 
kinematics, where ten quasars at $z>5.7$ are found to have rotation-like
velocity fields. Among these ten quasars, three have a significantly misaligned 
flux major axis and velocity gradient. This result suggests that complicated kinematics
are common in high-redshift quasars and that high-resolution observations are crucial 
to understanding high-redshift quasar host galaxies.

%%%%%%%%%%%%%%%%%%%%%%%%%%%%%%%%%%%
%%%%%%%%%%%%%%%%%%%%%%%%%%%%%%%%%%%
%  Section 5. Discussion          %
%%%%%%%%%%%%%%%%%%%%%%%%%%%%%%%%%%%
%%%%%%%%%%%%%%%%%%%%%%%%%%%%%%%%%%%

\section{A Maximum Starburst System With Oversized Black Hole at Cosmic Dawn}\label{sec:discuss}

\subsection{A Maximum Star Forming Rotating System} \label{subsec:sfr}

Our analysis shows that the host galaxy of J0439+1634 
is a compact ULIRG with vigorous star formation.
Assuming that the dust continuum traces the SFR surface density (SFRD),
we estimate the SFRD within the continuum half-light radius
%$R_\text{eff,cont}=0.76 \text{ kpc}$
to be $\Sigma_\text{SFR}\approx800 M_\odot\text{ year}^{-1}\text{ kpc}^{-2}$.
Such a high $\Sigma_\text{SFR}$ 
is close to the highest SFRD values seen in the universe
\citep[$\sim10^3 M_\odot\text{ year}^{-1}\text{kpc}^{-2}$; e.g.,][]{walter09}
and approaches the Eddington-limit of star formation \citep{thompson05}.
In addition, we estimate the maximum SFR proposed by \citet{elmergreen99},
where the gas is assumed to collapse on a free-fall timescale,
$t_\text{ff}=\sqrt{2R^3/GM}=1.06\sin(i)\times10^7$ year. 
The maximum possible SFR is $\epsilon M_\text{gas}/t_\text{ff}$,
where $\epsilon$ is the efficiency of gas turning into stars.
This argument suggests that J0439+1634 has a maximum possible SFR of
$\epsilon\sin^{-1}(i) \times 1.12\times 10^3 M_\odot \text{ year}^{-1}$ 
within a radius of $R<2 \text{ kpc}$.
For any reasonable inclination angle $(i\gtrsim5^\circ)$,
a high star formation efficiency is required $(\epsilon\gtrsim0.1)$.
Our analysis suggests that
J0439+1634 is forming stars at the maximum possible rate.

The rich gas reservoir and the vigorous star formation 
of J0439+1634 could be a remnant of a recent major merger or
strong cold gas inflow \citep[e.g.,][]{dekel09}. 
%Given the regular shape of the dust and {\cii} emission and the 
%smooth structures in the position-velocity plot,
%if J0439+1634 is a major merger, the merging process must have ended.
The major merger remnant scenario is promising because it provides a 
natural explanation to the misalignment between the major axis and the velocity gradient,
i.e., the star-formation regions are not yet evenly distributed in the rotating disk.
Under the current resolution, small-scale structures will get smoothed out,
and the flux distribution mimics a S\'ersic profile.
Upcoming high-resolution ALMA data could reveal these possible structures.

%The maximum star formation of J0439+1634 thus put questions on
%where the 

%The galaxy is in a regular shape and is rotation-dominated,
%with no evidence of significant outflows and/or signs of mergers.

%The host galaxy of J0439+1634 is a compact ULIRG
%with an SFR of.

\subsection{SMBH-Host Co-evolution} \label{subsec:coev}

\citet{fan19} measures the SMBH mass of J0439+1634 to be 
$M_\text{BH}=(4.29\pm 0.60)\times10^8M_\odot$,
which gives $M_\text{BH}/M_\text{dyn}=0.055\sin^2(i)$.
Assuming J0439+1634 follows the local relation in \citet{kh13} yields
$M_\text{BH}/M_\text{host}=0.005$.
A face-on rotation model with inclination $i=17^{\circ}$
 moves J0439+1634 onto the local $M_\text{BH}-M_\text{host}$ relation, and
a fiducial inclination angle of $i=60^\circ$
yields $M_\text{BH}/M_\text{host}=0.04$.
This result is similar to that in many high-redshift quasars
\citep[e.g.,][]{venemans17, decarli18, wang19b},
which have $M_\text{BH}/M_\text{host}$ several times
higher than the local relation.
 
With the SMBH mass and the observed central velocity dispersion,
we estimate the size of the SMBH's sphere of influence:
\begin{equation}
    r_h=\frac{GM_\text{BH}}{\sigma^2}.
\end{equation}
Both the image-plane moment 2 (Figure \ref{fig:clean})
and the reconstructed source-plane moment 2 (Figure \ref{fig:reconstruct})
show roughly constant velocity dispersion across the galaxy.
We thus adopt the median value of the source-plane moment 2 map, 
$\sigma_\text{med}=94 \text{ km s}^{-1}$,
which gives $r_h = 0.18 \text{ kpc}$.
The most extended configuration of ALMA delivers 
a resolution of $\sim0\farcs02$. For the region near the 
SMBH, we apply the magnification of the optical quasar from \citet{fan19},
$\mu_\text{quasar}=51.3$. An image-plane resolution of $0\farcs02$
thus corresponds to a source-plane resolution of 
$\sim2.8\text{ mas}$ $(\sim 16 \text{ pc})$.
Thus, high-resolution ALMA observations will allow us to sample the 
SMBH's sphere of influence well
and to measure the mass of the SMBH directly via gas kinematics.

Direct measurement of SMBH mass has been possible only at low-redshift
\citep[for a review, see][]{kh13}.
For most high-redshift quasars, SMBH masses are measured 
based on the empirical relation between the continuum luminosity and 
the broad line region size \citep[e.g.,][]{vp06}, 
which has only been calibrated at $z\lesssim2$.
%The extreme magnification for the optical quasar of
J0439+1634 thus provides a unique opportunity to calibrate 
the SMBH mass measurement at high redshift.

\section{Conclusions} \label{sec:conclude}
We present ALMA observations of a gravitationally lensed quasar
at $z=6.52$, J0439+1634. We model the dust-continuum, 
the {\cii} emission, and the velocity field of the host galaxy. 
Our main conclusions are:

\begin{enumerate}
    \item The ALMA observations
    demonstrate that the three-image fiducial model in \citet{fan19} based on {\it HST} observations of the quasar is correct,
    ruling out the alternative models considered in \citet{fan19} .
    The default lensing model gives an overall magnification of 
    $4.53\pm0.05$ and $3.44\pm0.05$ for the continuum and {\cii} emission of the host galaxy, respectively.
    The average source-plane resolution is $\sim0\farcs15$ $(\sim0.8 \text{ kpc})$.
    %The host galaxy of J0439+1634 
    %The half-light radius of the continuum emission is
    %$0\farcs136$ or $0.76 \text{ kpc}$.
    %The positions of the dust continuum, 
    %the {\cii} emission, and the optical quasar are consistent.
    \item J0439+1634 is a compact ULIRG well-described by
    a compact S\'ersic profile. The S\'ersic index is close to one
    for both the continuum and {\cii} emission.
    The resolved regions in J0439+1634 follow the ``{\cii} deficit,"
    suggesting that the deficit is related to the sub-kpc
    properties of the ISM.
    \item J0439+1634 has a rotation-like velocity field, but it cannot be well described as 
    an axisymmetric rotating thin disk.
    The maximum line-of-sight rotation velocity is $v_\text{max}\sin(i)=130\text{ km s}^{-1}$,
    with the inclination angle $i$ unconstrained.
    The dynamical mass within $2\text{ kpc}$ is 
    $7.9\times10^{9}\sin^{-2}(i)M_\odot$. 
    J0439+1634 is likely a gas-rich galaxy with a high gas-mass fraction.
    \item J0439+1634 is forming stars at the maximum possible rate.
    The star-formation rate surface density of J0439+1634 
    approaches the largest value seen in the universe and the Eddington limit.
    \item The SMBH-to-dynamical mass ratio of J0439+1634 is $0.055 \times \sin^2(i)$, 
    which suggests that J0439+1634 is likely to host an oversized SMBH
    compared to local relations.
    The size of the sphere of influence is $0.18 \text{ kpc}$.
    The most extended configuration of ALMA will resolve 
    the sphere of influence and allow us to measure the SMBH mass directly 
    using the gas kinematics.
\end{enumerate}

%Gravitational lensing enables 

Our lensing model incorporates the major features of J0439+1634 detected under low resolution ALMA data.
Future high-resolution ALMA observations with higher-resolution,
combined with pixelized lens modeling,
will reveal more detailed structures in the foreground lens
and quasar host galaxy.
Specifically, as discussed in Section \ref{subsec:coev},
the resolution around the SMBH will reach $\sim16 \text{ pc}$ within the SMBH sphere of influence.
The power of gravitational lensing makes J0439+1634 
a valuable object for a case study, which 
 will provide crucial and previously inaccessible information
about the coevolution of SMBHs and their hosts at $z>6$.

%Specifically, the optical quasar has a magnification of $\mu=51.4$
\acknowledgments
This paper makes use of the following ALMA data: ADS/JAO.ALMA\#2018.1.00566.S. ALMA is a partnership of ESO (representing its member states), NSF (USA) and NINS (Japan), together with NRC (Canada), MOST and ASIAA (Taiwan), and KASI (Republic of Korea), in co- operation with the Republic of Chile. The Joint ALMA Observatory is operated by ESO, AUI/NRAO and NAOJ.
The National Radio Astronomy Observatory (NRAO) is a facility of the National Science Foundation operated under cooperative agreement by Associated Universities, Inc.
MY, JY and XF acknowledges supports by NSF grants AST 15-15115 and AST 19-08284. MY acknowledges support from NRAO Student Observing Support (SOS) award SOSPA6-002.
FW thanks the support provided by NASA through the NASA Hubble Fellowship grant \#HST-HF2-51448.001-A awarded by the Space Telescope Science Institute, which is operated by the Association of Universities for Research in Astronomy, Incorporated, under NASA contract NAS5-26555.
JS acknowledge the support provided by NASA through the NASA Hubble Fellowship grants  \#HST-HF2-51446 awarded by the Space Telescope Science Institute.
KCL acknowledges support from the US NSF under grants AST-1715213 and AST-1716127 and from the US NSF NRAO under grant SOSPA5-001.
RW acknowledges the supports from the National Science Foundation of China (NSFC) grants No.11721303, 11991052,
and the National Key Program for Science and Technology Research and Development (grant 2016YFA0400703).
CK acknowledges support from US NSF grant AST-1716585.
BPV acknowledges funding through the ERC Advanced Grant 740246 (Cosmic Gas).
\facilities{ALMA}
\software{VISILENS}


\bibliography{sample63}{}
\bibliographystyle{aasjournal}



%\tablenum{1}
%\tablecaption{Lens model parameters}
%\label{tbl:lensparams}
%\tablewidth{0pt}
%\tablehead{
%\colhead{} & \colhead{Redshift} & \colhead{Mass} & \colhead{$\Delta$RA} &
%\colhead{$\Delta$Dec} & \colhead{$b/a$} & \colhead{PA} & \colhead{$n$} & 
%\colhead{Flux} & \colhead{$\mu$}\\
%\colhead{} & \colhead{} & \colhead{$(M_{\odot})$} & \colhead{(arcsec)} &
%%\colhead{(arcsec)} & \colhead{} & \colhead{(deg)} & \colhead{} & \colhead{}
%& \colhead{}
%}
%%\decimalcolnumbers
%\startdata
%{Lens} & [0.67] & [$2.071\times10^{10}$] & $0.416\pm0.002$ & $-0.073\pm0.002$ & [0.35] & [103.1] & - & - & - \\
%{Continuum} & [6.5187] & - & $0.227\pm0.003$ & $-0.023\pm0.003$ & $0.571\pm0.013$ & $7.8\pm1.0$ & $1.56\pm0.05$ & $3.008\pm0.051\text{ mJy}$ & $5.180\pm0.077$\\
%{{\cii}} & [6.5187] & - & 0 & 0 & 0 & 0 & 0 & 0 & 0\\
%\enddata
%\tablecomments{}
%\end{deluxetable*}

\end{document}